\documentclass[pra,twocolumn,showpacs,preprintnumbers,
amsmath,amssymb,tightenlines,epsfig]{revtex4}

\usepackage{bm}
\usepackage{graphicx}

\newcommand{\unit}{\hat{\bf n}}

\newcommand{\rv}{{\bf r}}

\newcommand{\beq}{\begin{equation}}
\newcommand{\eeq}{\end{equation}}
\newcommand{\bea}{\begin{eqnarray}}
\newcommand{\eea}{\end{eqnarray}}
\newcommand{\up}{\uparrow}
\newcommand{\down}{\downarrow}

\renewcommand{\>}{\rangle}
\renewcommand{\(}{\left(}
\renewcommand{\)}{\right)}

\newcommand{\commentout}[1]{{}}
\newcommand{\half}{\hbox{$1\over2$}}

\begin{document}

\title{Stable particlelike solitons with multiply-quantized vortex lines\\ in Bose-Einstein condensates}
\author{J. Ruostekoski}
\affiliation{Department of Physics, Astronomy and Mathematics, University of
Hertfordshire, Hatfield, Herts, AL10 9AB, UK}

\begin{abstract}
We show that a multiply-quantized vortex line core can be energetically stable in a harmonically trapped two-component atomic Bose-Einstein condensate. The structure, in which the condensate component with the vortex line is surrounded by the second component forming a vortex ring, can be identified as a particlelike Skyrmion with a multiply-quantized winding number, and closely resembles cosmic vortons. 
\end{abstract}

\date{\today}
\maketitle

\begin{figure}[!b]\vspace{-0.6cm}
\includegraphics[width=0.85\columnwidth]{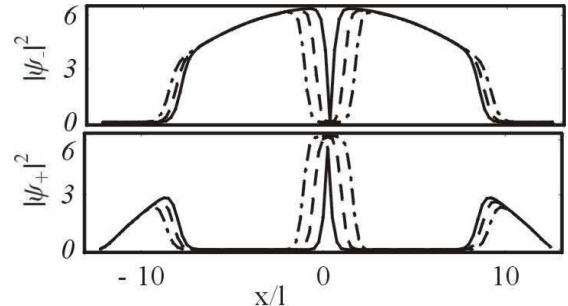}
\vspace{-0.3cm}
\caption{Density profiles for energetically stable multiply-quantized vortex cores in trapped Skyrmions. We display the 1D densities of the two BEC components $\psi_-$ (top) and $\psi_+$ (bottom) along the $x$ axis (in units of $10^{-4}l^{-3}$). We show the vortex line core of $\psi_-$ with the winding number one (solid line), two (dashed line), and three (dashed-dotted line) with the corresponding density profiles of $\psi_+$ exhibiting a singly-quantized vortex ring on the $xy$ plane. The size of the vortex line core region is approximately $2l$, $3l$, $4.5l$, respectively. }
\label{density}
\end{figure}

Among the most remarkable experiments in atomic Bose-Einstein condensates (BECs) are the studies of topological defects, such as quantized vortex lines and vortex lattices \cite{ANG}. A quantized vortex in a single-component BEC exhibits a line singularity of a 1D order parameter field. However, truly 3D topological objects can be afforded by multi-component BECs. It was recently shown that a localized particlelike soliton, or a Skyrmion, can be energetically stable in a trapped two-species $^{87}$Rb BEC under realistic experimental conditions \cite{SAV03}. In this paper we demonstrate that also multiply-quantized Skyrmions can be stable in a two-component BEC when the interparticle interaction between the two species is sufficiently strong. In the stable configurations one of the BEC components forms a multiply-quantized vortex line and is confined in a toroidal region inside the second component forming a vortex ring; see Fig.~\ref{density}. The second component provides the stability of the multiply-quantized vortex line core against splitting, and no external quartic potential is required, as in the case of multiply-quantized vortices in a single-component BEC \cite{ENG03}. Moreover, the centrifugal energy, associated with the circulation of the vortex line, prevents the shrinking of the vortex ring to zero size. As first noted in Ref.~\cite{BAT02}, the structure is closely related to cosmic vortons which are superconducting cosmic strings and may have been formed in the early Universe phase transitions \cite{cos}. Also nuclear matter in neutron stars can have a similar description \cite{BUC04}. The analogies provide an interesting possibility for cosmology in laboratory experiments using atomic BECs, as performed in superfluid liquid helium systems~\cite{BAU96,VOL03}.

Multi-component BECs can be prepared in magnetic traps by simultaneously confining different atomic species in the same trap or in optical dipole traps where the spin dynamics is no longer constrained by magnetic fields \cite{Pethick}. Here we consider two BEC components in perfectly overlapping isotropic magnetic trapping potentials. Such a system has been experimentally realized using the $|\up\>\equiv |F=2, m_f =1 \rangle$ and $|\down\>\equiv |1,-1 \rangle$ hyperfine spin states of $^{87}$Rb. In that case the inter- ($a_{\up\down}$) and intraspecies ($a_{\up\up}$ and $a_{\down\down}$) interaction strengths are nearly equal, with $a_{\down\down}:a_{\up\down}:a_{\up\up}::1.03:1:0.97$ \cite{HALL98}, and the order parameter of the two-component BEC has an approximate SU(2) symmetry. Since the scattering lengths satisfy $a_{\up\down}^2\agt a_{\up\up}a_{\down\down}$, the two species experience dynamical phase separation and can strongly repel each other \cite{HALL98}. The interatomic interactions of the two $^{87}$Rb components do not mix the atom population and the atom numbers of the both species are separately conserved.

The properties of the particlelike Skyrmions in a two-component atomic BEC have been addressed by several authors \cite{SAV03,Stoof,Stoof2,RUO01,BAT02,BAT02b,MET03}. An analogous configuration may also be formed in a spin-1 BEC \cite{Stoof2}. Other theoretical studies of spin defects and textures in multi-component BECs include, e.g., the structure of pointlike defects \cite{STO01,RUO03,SAV03b} and singular and coreless vortices \cite{SAV03b,YIP99,MIZ02,MUE,REI04,MUE04}. Vortices have been experimentally studied in two-component BECs \cite{MAT99,AND01,DUT01} and a coreless vortex has recently been engineered in an optically trapped $^{23}$Na spin-1 BEC experiment \cite{LEA02}. Moreover, topological defects in a two-component fermionic atomic gas in an optical lattice can result in particle number fractionilization \cite{RUO02}.

A Skyrmion is a localized particlelike soliton such that the order parameter field becomes uniform, with a constant asymptotic value, at sufficiently large distances from the particle \cite{SKY61}. As a result, we may enclose the Skyrmion by a sphere on which the field configuration has the constant asymptotic value; with the whole surface of the sphere identified with a single point. Then the physical space inside the sphere can be considered as a 3-sphere $S^3$ (a sphere in 4D). Moreover, the order parameter of a non-singular two-species BEC takes values on $S^3$ [or on SU(2)], and we may assign a winding number for the Skyrmion to characterize the mappings from the physical space to the order parameter space: $S^3\rightarrow S^3$.

For the soliton solution we parametrize the order parameter space $S^3$ by the unit vector $\unit=[-{\rm Im}(\psi_-),{\rm Re}(\psi_-),-{\rm Im}(\psi_+),{\rm Re}(\psi_+)]/\sqrt{\rho}$, in terms of the real and the imaginary parts of the two condensate wave functions $\psi_+(\rv)$ and $\psi_-(\rv)$. It is assumed that the total density $\rho(\rv)\equiv |\psi_+(\rv)|^2+|\psi_-(\rv)|^2$ is nonvanishing inside the sphere enclosing the Skyrmion. We then obtain the nonsingular configuration:
\beq\label{skyrmsol}
\begin{pmatrix}
\psi_+(\rv) \\ \psi_-(\rv)
\end{pmatrix}
= \sqrt{\rho}
\begin{pmatrix}
\cos\lambda-i\sin\lambda\cos\beta \\
-i\sin\lambda\sin\beta\exp(i\gamma)
\end{pmatrix} \,,
\eeq
where $(\lambda,\beta,\gamma)$ denote the spherical angles in 4D, such that $\unit=(\sin\lambda\sin\beta\cos\gamma,$ $\sin\lambda\sin\beta\sin\gamma,$ $\sin\lambda\cos\beta,$ $ \cos\lambda)$. Moreover, the winding number in terms of the spherical angles $\vec\alpha\equiv(\lambda,\beta,\gamma)$ reads:
\beq
\label{win1}
W={1\over 2\pi^2}\int d^3 x\, \sin^2\lambda\sin\beta\,
{\rm Det}\({\partial \alpha_i\over\partial x_j}\)\,.
\eeq
The integration is over the volume of the sphere enclosing the soliton, where $\rho$ is assumed to be nonvanishing, and $x_j$ denote the Cartesian coordinates. 
\begin{figure}[!b]\vspace{-0.5cm}
\includegraphics[width=0.18\columnwidth]{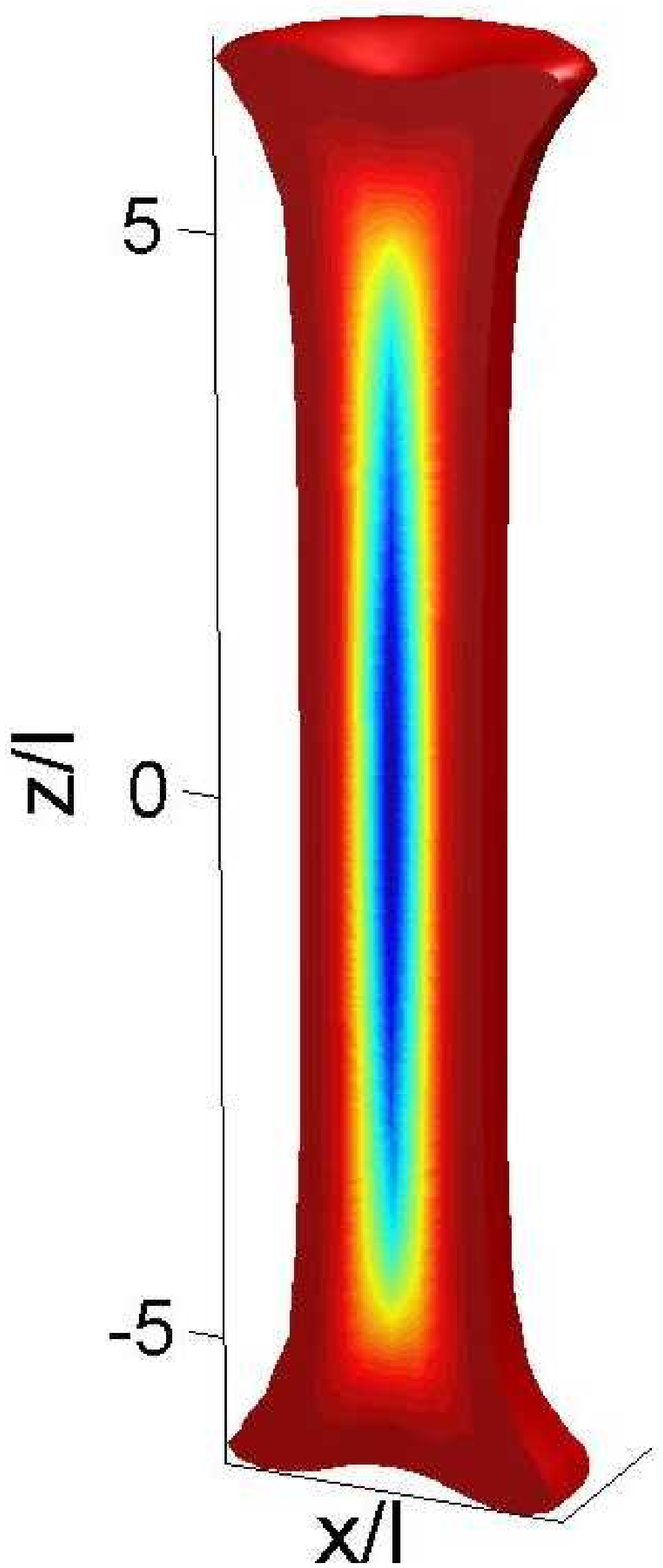}
\includegraphics[width=0.29\columnwidth]{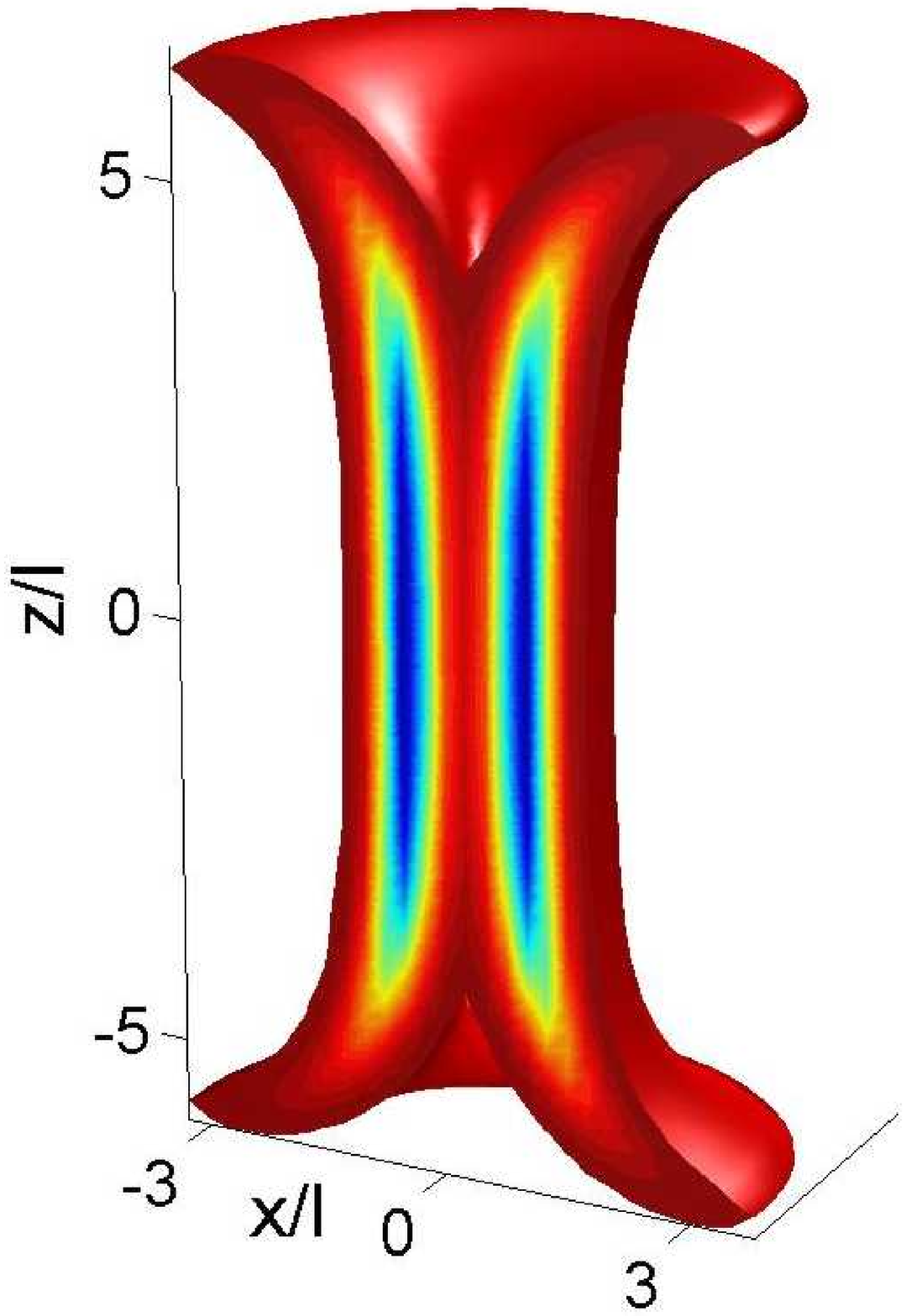}
\includegraphics[width= 0.36\columnwidth]{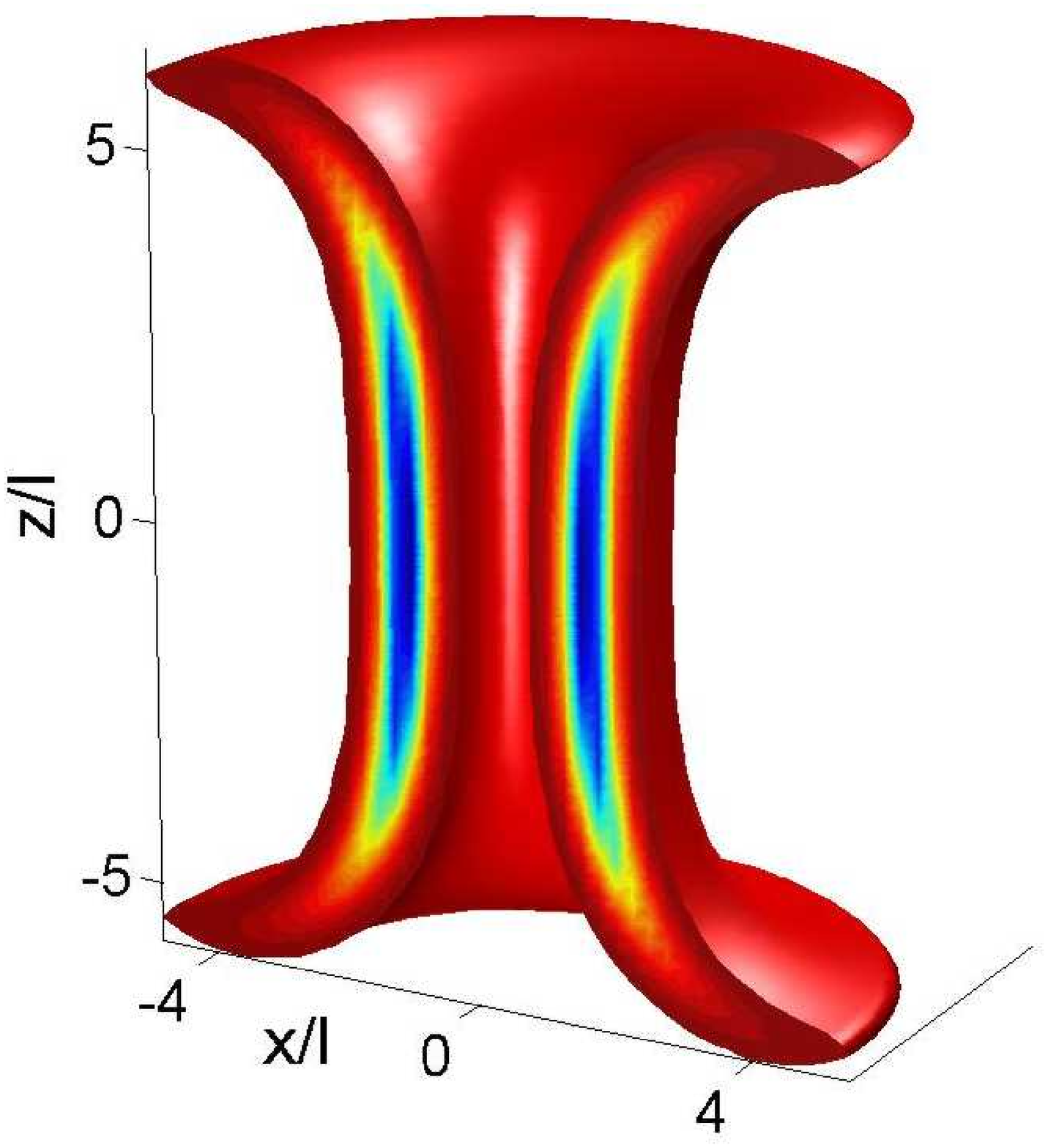}
\vspace{-0.3cm}
\caption{Charge densities for energetically stable trapped Skyrmions with $W=1,2,3$ (from left to right), for $y>0$. On the $xz$ plane the charge density between isosurface sections ranges from ${\cal W}=0.001$ (in units of $l^{-3}$) at the boundary (red) to ${\cal W}=0.017$ (left), ${\cal W}=0.075$ (middle), ${\cal W}=0.055$ (right) at the center (purple or dark grey). The charge is concentrated close to the interface of the two species.}
\label{isoplot}
\end{figure}

We numerically investigated the energetic stability of the BEC wave functions for the initial states which have different values of the winding number $W$. Although generally $\lambda=\lambda(\rv)$ is anisotropic, as an initial state we may use the spherically symmetric solution $\lambda=\lambda(r)$, with $\lambda(0)=0$, since the minimization of the energy does not preserve the spherical symmetry of the initial state. It is also illuminating to study the general structure of the solutions using this spherically symmetric ansatz. By inserting $\beta=k\theta$ and $\gamma=m\phi$ in Eq.~(\ref{win1}), with $k$ and $m$ integers and ($\theta,\phi$) the spherical angles, we obtain, for the even values of $k$, $W=0$, and for the odd values of $k$:
\beq
W= {m\over\pi} \left\{ \lambda(r_0)-\half \sin[2\lambda(r_0)]\right\}\,,
\eeq
where $r_0$ is the radius of the sphere enclosing the soliton. The winding number has an integer value $W=nm$, for $\lambda(r_0)=n\pi$, with $n$ integer. For $\lambda(r_0)=(n+1/2)\pi$, we obtain `hedgehog' solutions with $W=(n+1/2)m$. 

The Skyrmion solution (\ref{skyrmsol}), with $\beta=\theta$, $\gamma=m\phi$, and $\lambda(0)=0$ and $\lambda=n\pi$ at large distances from the particle, represents a vortex line, with the winding number $m$, in the component $\psi_-$ (the ``line component") and $n$ vortex rings in $\psi_+$ (the ``ring component"). The vortex rings are formed on the circles $z=0$, $r=\lambda^{-1}[(n-1/2)\pi]$. The line component, with the vortex line oriented along the $z$ axis, is confined inside the ring component and fills the toroidal core regions of the vortex rings. The winding number $W$ is topologically invariant if the asymptotic boundary values are not altered and $\rho$ remains nonvanishing for $r<r_0$. We may also obtain a solution for $W=nm$, e.g., by having a single vortex ring in $\psi_+$, with the winding number $n$, threaded by $m$ singly-quantized vortex lines in $\psi_-$, as long as $\psi_-$ with $m$ units of circulation is confined inside the circulating $\psi_+$. By assuming a two-species BEC to be an incompressible gas with a constant total atom density in the homogeneous space, the stable $W=2$ Skyrmions were obtained in Ref.~\cite{BAT02}. We show that the multiply-quantized vortex cores in a two-species BEC tend to split due to the inhomogeneous density, similarly to the vortices in a single-component BEC \cite{ENG03}. Nevertheless, we demonstrate it is still possible to find stable configurations.

The energetic stability of the Skyrmion in an inhomogeneous finite-sized system is a highly nontrivial concept. The configuration may collapse, the asymptotic boundary conditions of the trapped BEC may be violated, or the structure may not remain as singularity free.
Generally, additional physical mechanisms, such as rotation or optical potentials, will be required for stability \cite{SAV03}. Therefore, it is remarkable that energetically stable particlelike solitons can exist under experimental conditions of $^{87}$Rb, even in the presence of a strong external confinement due to the harmonic trap \cite{SAV03}.

Even in an infinite, homogeneous system topological particlelike solitons can be unstable against shrinking to zero size. If the energy does not contain any higher order gradient terms than quadratic, a simple spatial rescaling of the wave functions shows that the energy decreases monotonically with the size of the soliton, in the absence of additional stabilizing features. However, in a two-component BEC, without internal spin dynamics, the number of atoms in the two species is separately conserved, which can stabilize the Skyrmion against collapse to zero size \cite{BAT02}. In the phase separation regime the two species can strongly repel each other and the toroidal filling due to the line component provides a $1/r^2$ centrifugal barrier that can prevent the vortex ring from contracting to zero radius due to its tension. 

In Ref.~\cite{SAV03} it was numerically found that the phase separation of the two species is necessary, but not sufficient, condition for the stability of the Skyrmion with the winding number $W=1$ in a trapped BEC. Using the parameters of the $^{87}$Rb vortex experiments \cite{MAT99}, it was also required for the stability that the minimum number of atoms is several millions \cite{SAV03}. If, in addition, the two-component BEC was rotated at the speed in a narrow window close to 0.085$\omega$, where $\omega$ denotes the trap frequency, the Skyrmion was stable, without the vortex line drifting in the inhomogeneous trap towards the edge of the atomic cloud. Alternatively, if the BEC density profile remained cylindrically symmetric, the stability could be provided by rotating sufficiently fast only the vortex line component. The stable particlelike soliton studied in Ref.~\cite{SAV03} consisted of a singly-quantized vortex ring in $^{87}$Rb $|\down\>$ component, threaded by a singly-quantized vortex line in $|\up\>$. In this paper we show that also multiply-quantized Skyrmions can be energetically stable in a trapped BEC, providing a unique system with stable multiply-quantized vortex lines. 
\begin{figure}[b]\vspace{-0.5cm}
\includegraphics[width=0.29\columnwidth]{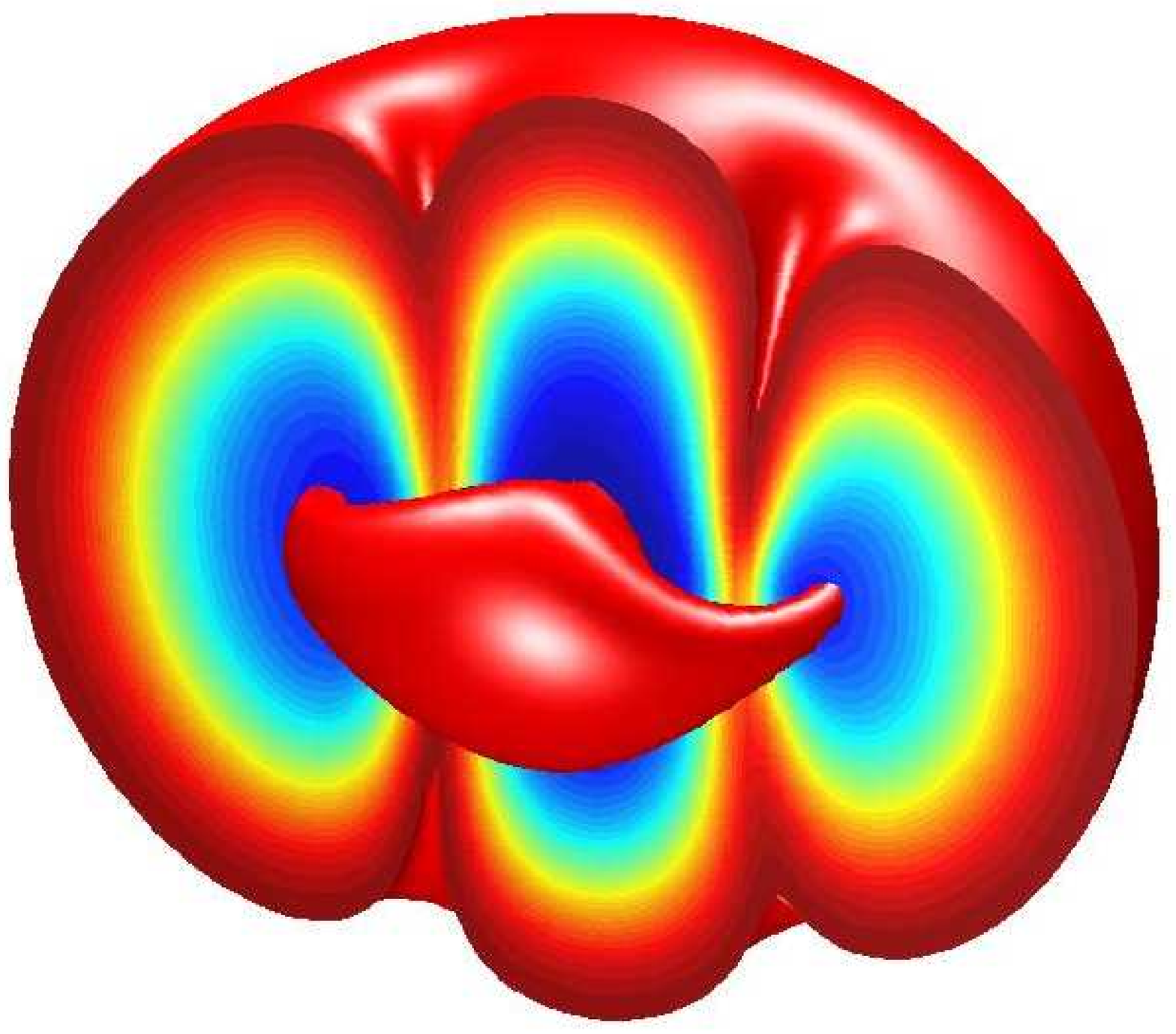}
\includegraphics[width=0.29\columnwidth]{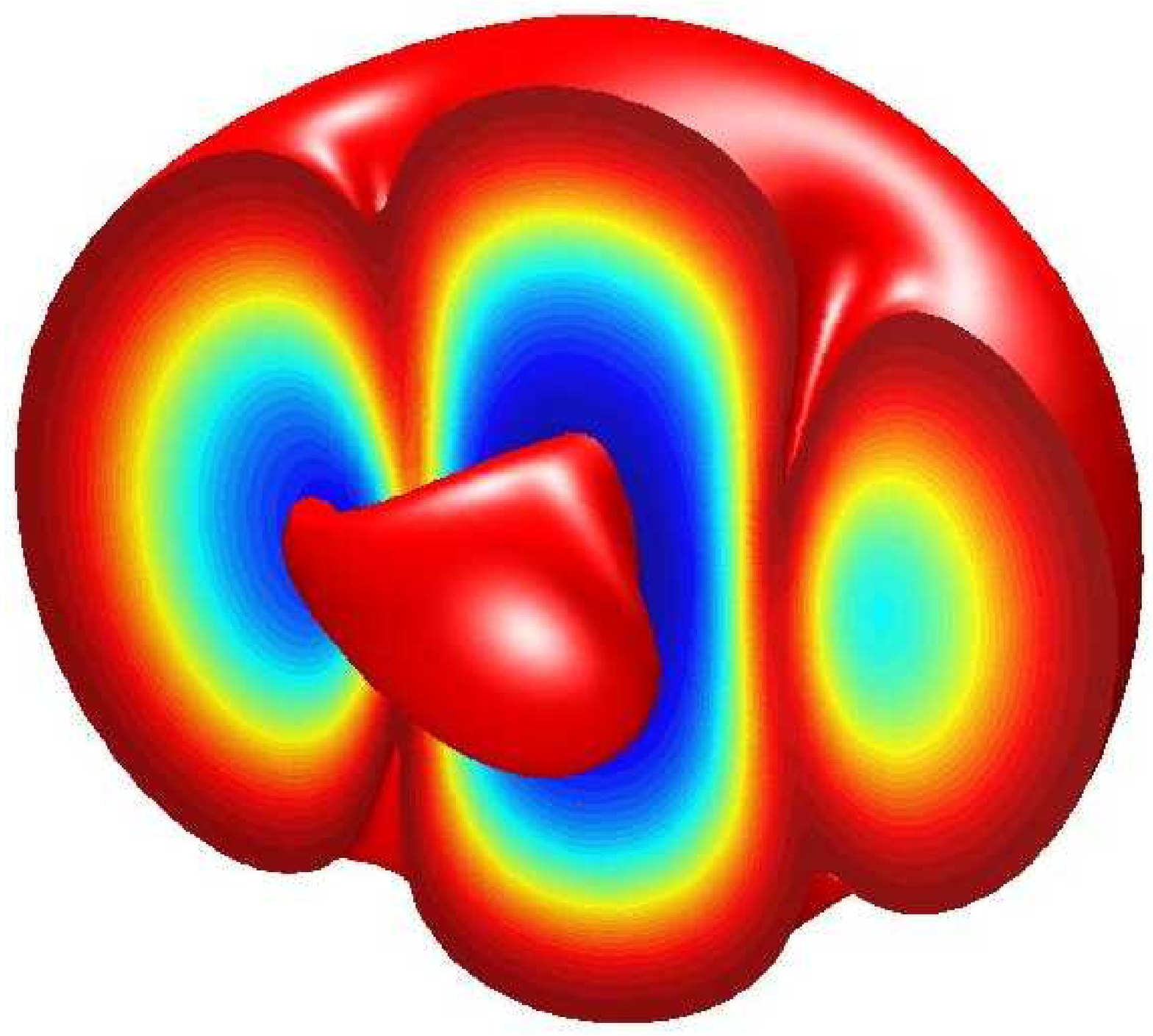}
\includegraphics[width= 0.29\columnwidth]{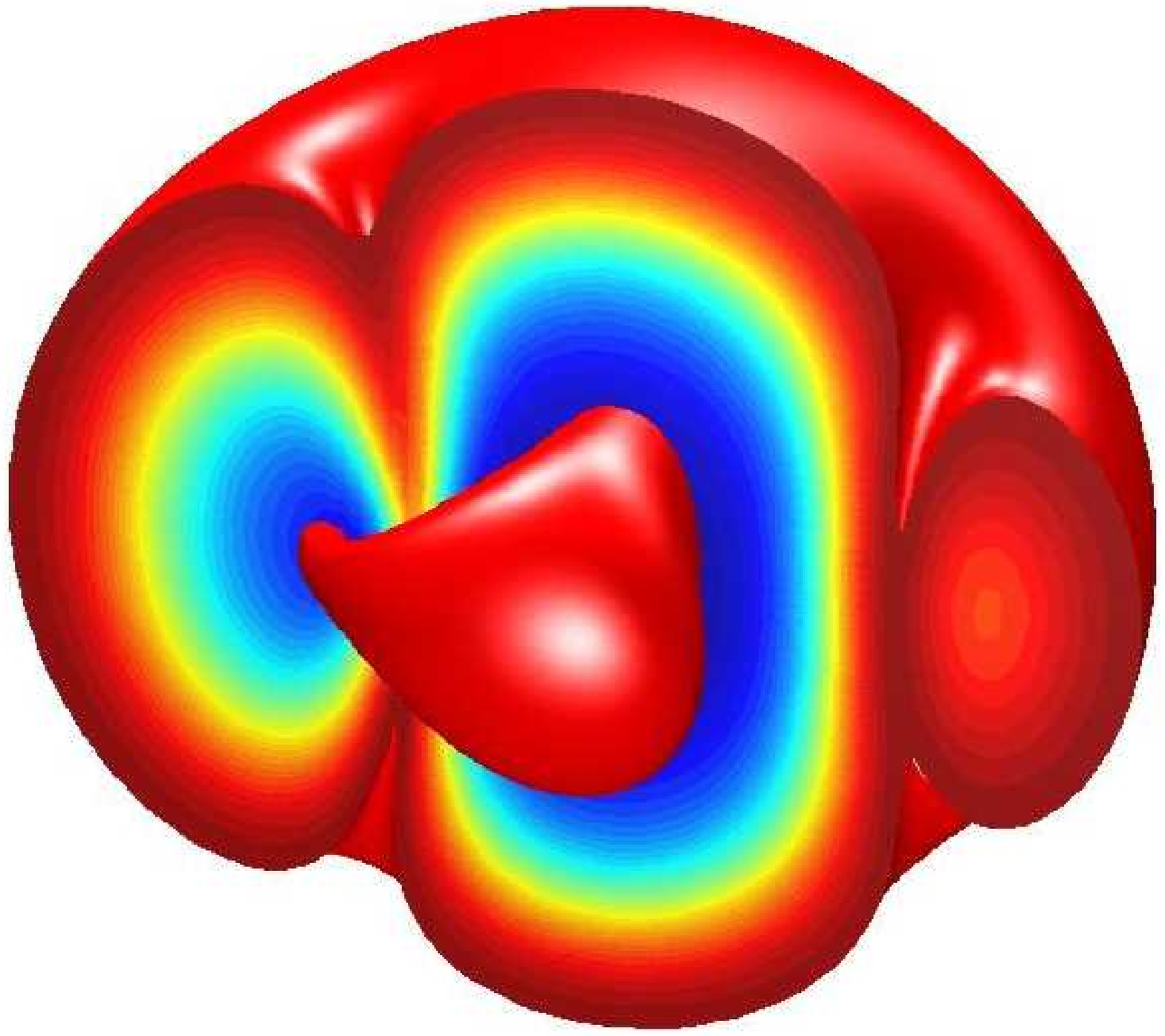}\vspace{-0.3cm}
\caption{The decay of a doubly-quantized vortex line core. The density profiles $|\psi_-|^2$ (for $y>0$) show the two singly-quantized vortices on the $xz$ plane between isosurface sections ranging from low at the boundary (red) to high at the center (purple or dark grey) after the double vortex has already split in the imaginary time evolution. On the left, the constant surface plot of $|\psi_+|^2$ (red torus) represents a vortex ring encircling the both vortex lines, and the Skyrmion winding number $W=2$. In the middle and on the right, at later times, one of the vortices has crossed the vortex ring which now only encircles one vortex line, and $W=1$. }
\label{split}
\end{figure}

We numerically studied the energetic stability of the particlelike solitons using the 3D nonlinear mean-field theory of the trapped two-component BEC. We minimized the energy of the Skyrmions by evolving the coupled Gross-Pitaevskii equations (GPEs),
\beq 
\label{gpe} 
i \hbar  \partial_t\psi_{i} = \big(H_0+
\sum_{j} \kappa_{ij} | \psi_j |^2 \big)
\psi_{i}\,, 
\eeq 
in imaginary time, for initial states exhibiting different winding numbers $W$, using the split-step method on a spatial grid of $128^3$. Here $H_0\equiv -\hbar^2\nabla^2/2m + V(\rv)$, with a perfectly overlapping isotropic trapping potential for both states $V(\rv)=m\omega^2 r^2/2$, and $\kappa_{ij} \equiv 4\pi\hbar^2 a_{ij} N /m$, where $N$ denotes the total atom number. We assume the harmonic oscillator length $l\equiv (\hbar/m\omega)^{1/2}=3.86 \mu$m, which for $^{87}$Rb corresponds to $\omega \simeq 2\pi \times 7.8$ Hz \cite{MAT99,com}. The winding number was calculated during the imaginary time evolution to determine the stability of the Skyrmion. We also numerically minimized the energy for the BECs rotating along the $z$ axis, in order to stabilize the Skyrmion against the drift towards the BEC boundary where the density is lower. This was done by evolving Eqs.~(\ref{gpe}) in imaginary time, in the rotating frame, obtained by replacing $H_0$ by $H_0-\Omega \hat{L}_z$, with $\hat{L}_z=i\hbar(y\partial_x-x\partial_y)$ and the rotation frequency denoted by $\Omega$. At every time step, we separately normalized the wavefunctions to $N_i/N$ to fix the atom number in each component. The simulations were fully 3D, without imposing any symmetry on the solution as it relaxed.

For particlelike solitons with multiply-quantized vortex lines in the line component, we used as an initial state Eq.~(\ref{skyrmsol}) with $W=nm$, for $n=1$ and for different integer values of $m$, embedded in a parabolic density profile. As shown before, we obtain this by choosing $\beta\rightarrow\theta$, $\gamma\rightarrow m\theta$, and a monotonic function $\lambda(r)$, with $\lambda(0)=0$ and $\lambda=\pi$ at the BEC boundary. In order to obtain stable Skyrmion configurations, $N$ had to be sufficiently large and the population difference between the two species $|N_+-N_-|$ sufficiently small. In the experimental values of $^{87}$Rb the slight difference between the scattering lengths $a_{++}:a_{+-}:a_{--}::1.03:1:0.97$, with $a_{+-}\simeq5.5$nm \cite{HALL98}, indicates that the SU(2) order parameter symmetry is only approximate, being broken down to U(1)$\times$U(1). As a result of the phase separation for $a_{+-}^2\agt a_{++}a_{--}$, the scattering length difference was previously found to provide a sufficient repulsion between the two species to stabilize the $W=1$ Skyrmion, when $N\agt8\times10^6$, for $N_+=N_-$ and $l=3.86 \mu$m \cite{SAV03,com}. However, here we were unable to find stable multiply-quantized vortex line cores for these parameters by considering $N$ up to $10^7$ atoms. For large $N$, when the Skyrmions did not collapse via shrinking, the multiply-quantized vortex cores for $m>1$ split into singly-quantized vortex lines which drifted towards the edge of the BEC. The winding number $W$ decayed by integer-multiples always when the vortex lines crossed the vortex ring core, by forcing the total density to zero at the crossing point and, therefore, by creating a singularity in the spin texture. In Fig.~\ref{split} we show the decay of a $W=2$ Skyrmion when the doubly-quantized vortex line has split into two singly-quantized vortices drifting to opposite directions. One of the vortices crosses the vortex ring core, which is numerically also confirmed by the sudden change in the value of the winding number from two to one. Here $N\simeq10^7$, $N_+=2N_-$, and $\Omega\simeq0.09/\omega$. Increasing the rotation speed did not stabilize the Skyrmions with $m>1$, since additional vortices entered the ring component due to faster rotation.

We found energetically stable multiply-quantized vortex line cores for $N< 10^7$ when we considered a stronger repulsion between the two BECs with $a_{+-}=a_{++}\simeq5.67$nm and $a_{--}\simeq5.34$nm. In Fig.~\ref{density} we show the stable density profiles of vortex lines with the winding numbers $m=1,2,3$, representing Skyrmions with $W=1,2,3$, for the case $N_+=N_-$, $N\simeq7.8\times10^{6}$, $a_{+-}\simeq5.89$nm, $a_{++}\simeq5.67$nm, and $a_{--}\simeq5.34$nm. Applying the rotation $\Omega\simeq0.09/\omega$ was found to be sufficient to stabilize the configurations at the trap center. The corresponding topological charge densities ${\cal W}(\rv)$, determined from
Eq.~(\ref{win1}) with $W=\int d^3 r {\cal W}(\rv)$, are displayed in Fig.~\ref{isoplot}. 

\begin{figure}[b]\vspace{-0.5cm}
\includegraphics[width=0.82\columnwidth]{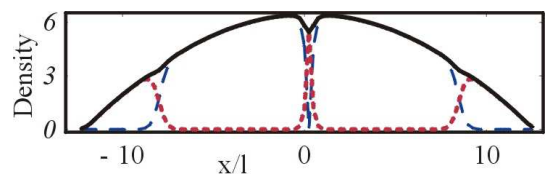}
\vspace{-0.3cm}
\caption{Density profiles for energetically stable $W=2$ particlelike solitons with two vortex rings. We show the 1D densities along the $x$ axis (in units of $10^{-4}l^{-3}$) for $\psi_-$ with a vortex line at $x=0$ (dashed blue line), $\psi_+$ (dotted red line) with two vortex rings, and the total density (solid line).}
\label{dble}
\end{figure}

It is also interesting to consider multiply-charged particlelike solitons with $n>1$ in the case of one singly-quantized vortex line ($m=1$). We found the stability to be very sensitive to the initial conditions. For two vortex rings on the $xy$ plane with slightly different radii, we found energetically stable $W=2$ configurations even using the scattering length values of the present $^{87}$Rb experiments \cite{HALL98}. In Fig.~\ref{dble} we show the stable case using the same parameters as in Fig.~\ref{density}. However, with one doubly-quantized vortex ring as an initial state, the double-ring eventually decayed to one singly-quantized ring, with the other ring pinching off at the center. This may explain why stable $W=2$ states with one vortex line were not found in the homogeneous space in Ref.~\cite{BAT02} where only a multiply-quantized vortex ring was considered.

The particlelike solitons could be created by a sequence of Raman pulses, with three orthogonal standing waves imprinting a vortex ring, as proposed in Ref.~\cite{RUO01}. A doubly-quantized vortex line has recently been experimentally prepared at the MIT by changing the magnetic field direction. Alternatively, multiply-quantized vortices could also be phase-imprinted as proposed in Ref.~\cite{RUO00}.

In conclusion, we showed how multiply-quantized vortex lines can be energetically stable, forming particlelike solitons with multiple winding numbers. The studied set of interaction parameter values could possibly be experimentally reached for different two-species atom mixtures, e.g., by using the Feshbach resonances. Moreover, we found the stable $W=2$ solitons for two vortex rings using the parameters of current $^{87}$Rb experiments. Although the stability requires reasonably large atom numbers, the stable configurations could possibly also be obtained using a smaller number of atoms, e.g., by first loading atoms to a weak trap and then adiabatically increasing the trapping frequencies to obtain the required nonlinearity \cite{com}. We did not study the stability of truly giant vortex line cores with $m\gg1$. This would be interesting, as giant vortex systems have also been created experimentally \cite{ENG03}.

This research was supported by the EPSRC.


\begin{thebibliography}{99}
\bibitem{com} In dimensionless units, $(\bar\rv,\bar{t})=(\rv/l,t\omega)$, the interaction coefficients $\bar\kappa_{ij} \equiv 4\pi N a_{ij} / l$, and the dynamics is unchanged in any scaling of length and time, which does not change $N/l$.
\end{thebibliography}
\end{document}